\documentclass{aastex62}

\usepackage{xspace}

\newcommand{\co}[2]{\ensuremath{^{#1}\textrm{C}^{#2}\textrm{O}}\xspace}
\newcommand{\car}[1]{\ensuremath{^{#1}\textrm{C}}\xspace}
\newcommand{\ox}[1]{\ensuremath{^{#1}\textrm{O}}\xspace}
\newcommand{\update}{} 

\received{Nov 9, 2018}
\submitjournal{ApJL}

\shorttitle{Isotopes in M dwarfs}
\shortauthors{Crossfield et al.}

\begin{document}

\title{Unusual Isotopic Abundances in a Fully-Convective Stellar Binary}

\correspondingauthor{I.J.M.\ Crossfield}
\email{iancross@mit.edu}

\author{I.J.M.\ Crossfield}
\affiliation{Department of Physics, and Kavli Institute for Astrophysics and Space Research, Massachusetts Institute of Technology, Cambridge, MA, USA}
\nocollaboration

\author{J.D.\ Lothringer}
\affiliation{Lunar and Planetary Laboratory, University of Arizona, Tucson, AZ, USA}
\nocollaboration

\author{B. Flores}
\affiliation{Department of Physics and Astronomy, California State University Northridge, Northridge CA, USA}
\affiliation{Department of Physics, and Kavli Institute for Astrophysics and Space Research, Massachusetts Institute of Technology, Cambridge, MA, USA}
\nocollaboration

\author{E.A.C.\ Mills}
\affiliation{ Department of Physics, Brandeis University, Waltham, MA, USA}
\nocollaboration

\author{R.\ Freedman}
\affiliation{NASA Ames Research Center, Moffett Field, CA, USA}
\affiliation{SETI Institute, Mountain View, CA, USA}
\nocollaboration

\author{J. Valverde}
\affiliation{Department of Physics, University of California, Santa Cruz, Santa Cruz, CA, USA}
\affiliation{Chabot-Las Positas Community College, Dublin, CA, USA}
\affiliation{Department of Physics, and Kavli Institute for Astrophysics and Space Research, Massachusetts Institute of Technology, Cambridge, MA, USA}
\nocollaboration

\author{B.\ Miles}
\affiliation{Department of Astronomy, University of California, Santa Cruz, Santa Cruz, CA, USA}
\nocollaboration

\author{X.\ Guo}
\affiliation{Department of Physics, and Kavli Institute for Astrophysics and Space Research, Massachusetts Institute of Technology, Cambridge, MA, USA}
\nocollaboration

\author{A.\ Skemer}
\affiliation{Department of Astronomy, University of California, Santa Cruz, Santa Cruz, CA, USA}
\nocollaboration



\begin{abstract}

Low-mass M dwarfs represent the most common outcome of star formation,
but their complex emergent spectra hinder detailed studies of their
composition and initial formation. The measurement of isotopic ratios
is a key tool that has been used to unlock the formation of our Solar
System, the Sun, and the nuclear processes within more massive stars.
We observed GJ~745AB, two M dwarfs orbiting in a wide binary, with the
IRTF/iSHELL spectrograph. Our spectroscopy of CO in these stars at the
4.7\,$\mu$m fundamental and 2.3\,$\mu$m first-overtone rovibrational
bandheads reveals \co{12}{16}, \co{13}{16}, and \co{12}{18} in their
photospheres. Since the stars are fully convective, the atomic
constituents of these isotopologues should be uniformly mixed
throughout the stars' interiors. We find that in these M dwarfs, both
\car{12}/\car{13} and \ox{16}/\ox{18} greatly exceed the Solar values.
These measurements cannot be explained solely by models of Galactic
chemical evolution, but require that the stars formed from an ISM
significantly enriched by material ejected from an exploding
core-collape supernova.  These isotopic measurements complement the
elemental abundances provided by large-scale spectroscopic surveys,
and open a new window onto studies of Galactic evolution, stellar
populations, and individual systems.
\end{abstract}



\section{Introduction \& Observations} \label{sec:intro}

Detailed analysis of the thermal emission spectra of stars smaller,
cooler, and lower-mass than the Sun is significantly more challenging
than for hotter, brighter stars.  These  M dwarfs
are relatively faint, emit most of their energy at wavelengths beyond
the visible, and their atmospheres are cool enough to contain numerous
molecules with many spectral features. Nonetheless these cool objects
are subjects of considerable study, both because they represent the
single most common outcome of star formation and because they appear
to be especially likely to host planetary systems
\citep{dressing:2015}.  By characterizing the chemical properties of M
dwarfs, we learn about the chemical enrichment and star formation
history of our own Milky Way galaxy, and hope to also learn about the
formation of planetary systems, including some of the best targets for
studying potentially habitable planets.

We used the iSHELL spectrograph \citep{rayner:2016} on the NASA
Infrared Telescope Facility to observe GJ~745A and B, two otherwise
indistinguishable M dwarfs with radii, masses, and metallicity all
roughly a third that of the Sun (see Table~\ref{tab:params}). Both
stars should be chemically homogeneous throughout, because they are
fully convective.  The stars lie just 0.05~mag in brightness, and
0.05~dex in luminosity, below a newly-identified gap in the lower main
sequence that separates fully-convective M-dwarfs from those that are
only partially convective \citep{jao:2018,macdonald:2018}.

We observed GJ~745A and B on the night of UT 2017-07-02 (Program
2017A110, PI Crossfield), acquiring $R=70,000$ (4.3 km~s$^{-1}$)
spectroscopy and mostly-continuous coverage from
4.52--5.24\,$\mu$m. The full details of our observations are listed in
Table~\ref{tab:params}. Conditions were photometric throughout the
night\footnote{As shown by the CFHT Skyprobe atmospheric transparency
  data archive,
  \url{http://www.cfht.hawaii.edu/cgi-bin/elixir/skyprobe.pl?plot&mcal_20170702.png}}.
We reduced the raw iSHELL data using the \texttt{SpeXTool} Data
Reduction package \citep{cushing:2004}.  \texttt{SpeXTool} flat-fields
raw images to correct for pixel-to-pixel variations and uses sky
emission lines in science frames for the wavelength calibration. The
calibrated M-band frames were then nod-subtracted (to remove sky
emission and hot pixels) and stacked to produce a set of master frames
for each star. After calibrating this master frame, spatial profiles
are computed, two one-dimensional spectra are extracted (one at each
nod position), and the two spectra are combined to produce a single
spectrum for each star.

We then correct for telluric absorption features by using the observed
A0V standard star (HR 7390), the science target star (GJ~745 A or B),
and a high-resolution model spectrum of Vega \citep{vacca:2004}. Since
A0V stars have spectra that are nearly featureless, the A0V spectrum
corrects the object spectra.  We also tune the depths and widths of
hydrogen absorption lines in the model to better match HR 7390 and
minimize residuals at these wavelengths. {\update Although HR~7390
  rotates more rapidly than Vega and is somewhat cooler, both of these
  factors are accounted for in the \texttt{SpeXTool} reduction: the
  former by convolving the Vega model spectrum with a rotational
  broadening kernel, and the second by adjusting the spectral slope
  based on the star's (B-V) color.} Finally, we remove parts of the
spectrum with obvious bad pixels and wherever S/N $<$ unity. {\update
  In practice, the choice of S/N cut-off is not especially significant
  since low-S/N parts of the spectrum are appropriately de-weighted
  when we calculate our weighted-mean line profile for each
  isotopologue.}

\section{Modeling}
\subsection{Stellar Spectra}
To measure the \car{12}/\car{13} isotopic ratio of GJ~745A and~B we
compare our observed spectra to {\update synthetic spectra generated
  from custom atmosphere models of the two stars, both spectra and
  models being derived from the PHOENIX atmosphere code}
\citep[Version 16,][]{husser:2013}. Our PHOENIX {\update model
  atmospheres contain} 64 vertical layers, spaced evenly in log-space
on an optical depth grid from $\tau = 10^{-10}-100$, spanning
$1.0-10^5$~nm. In our observed wavelength range, the models were
sampled at least every 0.01~nm. The models were run with H I, He I-II,
C I-IV, N I-IV, O I-IV, Mg I-III, and Fe I-IV in NLTE. We ran models
using the stellar parameters listed in Table~\ref{tab:params} for five
different \car{12}/\car{13} ratios --- 29.3, 89.9, 271.7, 908.1, and
2731.2, corresponding to \car{13} enrichments of 3x, 1x, 1/3x, 1/10x,
and 1/30x solar, respectively --- and three different \ox{16}/\ox{18}
ratios --- 165.3, 498.8, and 1497.7, corresponding to \ox{18}
enrichments of 3x, 1x, and 1/3x solar, respectively. We use a CO line
list \citep{goorvitch:1994} that contains lines for \co{12}{16},
\car{13}{16}, \co{12}{17}, \co{12}{18}, \co{13}{18}, \co{14}{16}, and
\co{13}{17}.

We verify our isotopic measurement by comparing a PHOENIX model of the
Sun to high-resolution spectra from Kitt Peak's Fourier-Transform
Spectrograph \citep{hase:2010}. Our Solar model gives an excellent
match to the known solar isotopic ratios. Including the atoms listed
above in NLTE in the Solar model changes the line depths of CO
isotopologues by 0.3\%, negligible compared to our current measurement
uncertainty.

\subsection{Measuring Isotopic Ratios}
\begin{figure}
\plotone{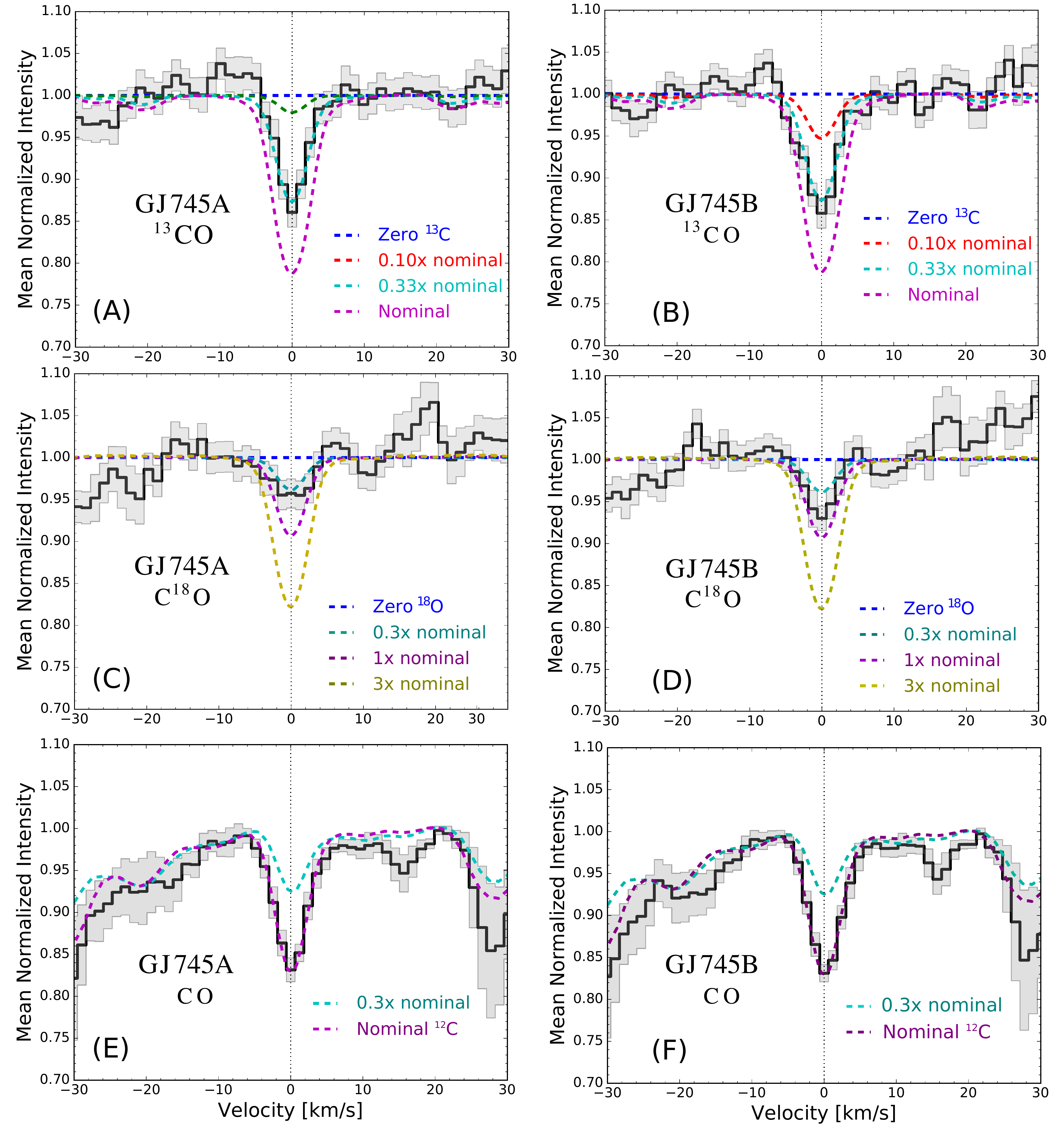}
\caption{Stacked absorption lines of CO isotopologues in GJ~745AB, showing clear evidence of both \co{13}{} (panels A and B), \co{}{18} (panels C and D), and the abundant \co{12}{16} (panels E and F).  The black curve and gray shaded region indicate our measurements and their 68.3\% confidence intervals; dashed lines indicate spectral models with the indicated abundances of the highlighted isotopes relative to the nominal abundance of [Fe/H]= -0.4 dex. \label{fig:lines}}
\end{figure}

The highest-S/N regions of our spectra are 4.6--4.7 micron, where
tellurics are relatively weak and stellar spectra are dominated by
\co{12}{16} lines.  We used the HITEMP database \citep{rothman:2010}
to identify \co{13}{} and \co{}{18} lines that are relatively clear of
tellurics and other strong absorption lines, as listed in
Table~\ref{tab:lines}.  Most of the \co{13}{} lines are individually
visible but all have fairly low statistical significance, while the
individual \co{}{18} lines can only barely be discerned by eye. To
boost our S/N we construct a single line profile by taking the
weighted mean of each line (after linearly continuum-normalizing each
line using the regions listed in Table~\ref{tab:lines}). The resulting
mean line profiles, shown in Figure~\ref{fig:lines}, clearly reveal
the strong signature of both \co{13}{} and the rarer \co{}{18} in both
GJ~745A and B.

We measure the \car{13}/H and \ox{18}/H abundance ratios for each star
by interpolating our grid of PHOENIX models so as to minimize the
$\chi^2$ calculated from the mean observed and modeled lines ({\update
  after removing a linear pseudocontinuum as described above}). We
infer 1$\sigma$ confidence intervals using the region where $\Delta
\chi^2$~$\le$ unity \citep{avni:1976}.  We find that the accessible
\co{12}{} lines in the fundamental band are too strongly saturated to
tightly constrain the stars' \car{12}/H abundances, so we instead use
weaker lines in the first overtone band from 2.1--2.5\,$\mu$m iSHELL
spectra of the binary taken on the same night.

{ \update To verify that using different CO lines in different bands does not
bias our results, we compare the individual intensities of the lines
used in our analysis from each of three sources: Goorvitch et
al. (\citeyear{goorvitch:1994}; still used in constructing our PHOENIX
model spectra; HITEMP \citep{rothman:2010}; and a custom list
constructed by Gordon et al.\ (R. Freedman, private communication). We
find clear evidence of systematic offsets in the ($gf$) values, with
consistent offsets for each combination of isotopologue and
linelist. These numbers imply that the inferred isotopic abundances
may suffer from systematic biases at the ~2\% level. Since we
currently measure $^{12}$C/$^{13}$C to only 10\% precision, this 2\%
effect does not significantly impact our current analysis. }

With this approach, we measure [C/H] = $-0.37$ and $-0.39$ dex for
GJ~745A and B, respectively -- entirely consistent with their iron
depletion of [Fe/H]= $-0.4$~dex.  Since CO is a poor measure of oxygen
abundance, we use the [C/H] and [O/H] abundances observed in M dwarfs
\citep{tsuji:2016, souto:2017, souto:2018} to infer \ox{16}/H. These
observations include stars with a range of metallicities and are
consistent with [O/H] = [C/H], so we assume this ratio also holds for
our targets.

We find that the best-fit abundance ratios vary by roughly 5\%
depending on which particular lines we use for stacking and which
range of velocities we use to calculate $\chi^2$. We therefore assume
an additional ($5\% \times \sqrt{2}$) = 7.1\% systematic uncertainty
for our isotopic measurements. Nonetheless our total uncertainties are
dominated by measurement noise.

\subsection{Discussion}
Our spectra clearly reveal multiple rare isotopologues of carbon
monoxide (CO).  \co{13}{} has been inferred from medium-resolution
2.3\,$\mu$m spectroscopy, but \co{}{18} has not been measured in any
dwarf stars beyond the Sun. For both \co{}{18} and \co{13}{}, we find
isotopic abundance ratios significantly discrepant from the Solar
values. Whereas the Sun has \car{12}/\car{13} = $93.5 \pm 3.1$ and
\ox{16}/\ox{18} = $525 \pm 21$ \citep{lyons:2018}, for GJ~745 A and B
we find \car{12}/\car{13} = $296 \pm 45$ and $224 \pm 26$, and
\ox{16}/\ox{18} = $1220 \pm 260$ and $1550 \pm 360$, respectively (see
Table~\ref{tab:params}).  The ratio of our \co{12}{}/\co{13}{} and
\co{}{16}/\co{}{18} abundance ratios gives \co{13}{}/\co{}{18}, a
quantity more accurately measured in many astronomical objects because
these two isotopologues are typically both optically thin (unlike
\co{12}{16}). We find \co{13}{}/\co{}{18}= $4.1 \pm 1.1$ and $6.9 \pm
1.8$ for GJ~745A and B respectively.

Although the individual isotopic ratios are nonsolar, our measured
\co{13}{}/\co{}{18} ratios are broadly consistent with the Solar value
of $5.6 \pm 0.2$ \citep{lyons:2018} and typical values for the
interstellar medium (ISM) in our Galaxy and the disks of other spiral
galaxies \citep[ratios of 5-10;][]{jimenez-donaire:2017}, and are also
consistent with values inferred for the ISM in the nuclei of starburst
and more quiescent galaxies on 10--100~pc scales \citep[ratios of
  2.5--4;][]{meier:2004}.  However, low \co{13}{}/\co{}{18} values
measured in the ISM of other galaxies are often coincident with low
\ox{16}/\ox{18} values attributed to abundant \ox{18}
\citep{meier:2004}, the opposite of what we observe for our stars. The
isotopic ratios of GJ~745AB are also consistent with young stellar
objects (YSOs) and ionized gas regions in our own Milky Way
\citep{smith:2015}, but the abundances of these Galactic objects are
still inconsistent with GJ~745AB since newly-forming stars should
have higher metallicity. One must also take care in comparing to ISM
values, as these can be affected by processes such as selective
photodissociation \citep{bally:1982} and fractionation
\citep{watson:1976} which can lead to the preferential formation or
destruction of certain isotopologues of a molecule, such that the
measured molecular abundances are not representative of the true
abundance of an isotope.

The individual isotopic abundances of our stars {\update cannot be
  matched} by standard models of Galactic chemical evolution
\citep{kobayashi:2011} despite the broad consistency between
\co{13}{}/\co{}{18} in GJ~745A and B and some Galactic
measurements. These chemical models predict much higher
\ox{16}/\ox{18} ratios for our observed \car{12}/\car{13} ratio --- or
equivalently, lower \car{12}/\car{13} ratios at the known stellar
metallicity. Some deviations are seen from predictions of the
evolution of \car{12}/\car{13} with time and from the observed trends
in isotope ratios with Galactic radius. The current \car{12}/\car{13}
in the Solar neighborhood ISM is 30\% smaller than that in the Sun
with little corresponding change in \ox{16}/\ox{18}
\citep{polehampton:2005,milam:2005}. There is also a factor of $\sim$2
dispersion in the present day \car{12}/\car{13} ratio in the Milky Way
ISM at a given Galactic radius \citep{milam:2005}, but this intrinsic
scatter is still too small to explain the carbon isotope ratios that
we see.

What, then, could cause such surprisingly high isotopic ratios? Different
astrophysical phenomena affect \car{12}/\car{13}, \ox{16}/\ox{18}, and
[Fe/H] in different ways. Accretion of gas enriched by mass loss
products from evolved, asymptotic giant branch stars has been
suggested to explain the Sun's oxygen isotope ratios
\citep{gaidos:2009}, but fails to match our observations since these
evolved stars have much lower \car{12}/\car{13} ratios than we see
\citep{sneden:1986}. Carbon-rich giant stars of the R Corona Borealis
type often have \car{12}/\car{13} $\ge$ 100 \citep{fujita:1977} and
undergo frequent mass loss \citep{clayton:1996}, but their
\ox{16}/\ox{18} ratios are lower than the Solar value
\citep{clayton:2005}, contrary to what we observe.  The relatively
large \car{12}/\car{13} ratios seen in ULIRGs have been suggested to be
due to infall of low-metallicity gas \citep{casoli:1992a}, as is seen
in the center of our Galaxy \citep{riquelme:2010}. However, such
pristine gas would have too little \ox{18} to reproduce the observed
\ox{16}/\ox{18} ratios in these stars. While a combination of both a
starburst (decreasing the \ox{16}/\ox{18}) and an infusion of low
metallicity gas (increasing the \ox{16}/\ox{18}) has been suggested to
simultaneously allow for somewhat similar isotopologue ratios
\citep[\co{12}{}/\co{13}{} $> 90$, \ox{16}/\ox{18} $>
  900$;][]{konig:2016}, such a scenario is quite complex.

Alternatively, higher isotopic ratios can be caused by the inclusion
of material from dramatic episodes of rapid nucleosynthesis
\citep{casoli:1992b}, such as accretion of supernova ejecta.  It is
this model that best explains our data. We use models of the evolution
of Galactic abundances to represent the initial ISM
\citep{kobayashi:2011} and model the ejecta composition using
simulated isotopic yields of CCSN \citep{woosley:1995}.  We construct
a model in which the free parameters are the initial [Fe/H] of the ISM
and SN progenitor star, the progenitor mass, and the fraction of
resulting stellar mass consisting of SN ejecta.

We find that enrichment by material from a CCSN with progenitor mass
of roughly $21 M_\odot$ {\bf is required to explain} our observations,
independent of the assumed Galactic environment model
\citep{kobayashi:2011} -- halo, thick disk, Solar neighborhood, or
bulge.  We can anyway exclude both the thick-disk model and the halo
model because GJ~745's 3D motion in the Galaxy ($U =
-45.8$~km~s$^{-1}$, $V = +17.3$~km~s$^{-1}$, $W = +22.2$~km~s$^{-1}$)
is inconsistent with the motion expected for thick-disk or halo stars
\citep{fuhrmann:2004}.  We also exclude the bulge model because
GJ~745AB is $<10$~pc away, far from the Milky Way's bulge.

\begin{figure}
\plotone{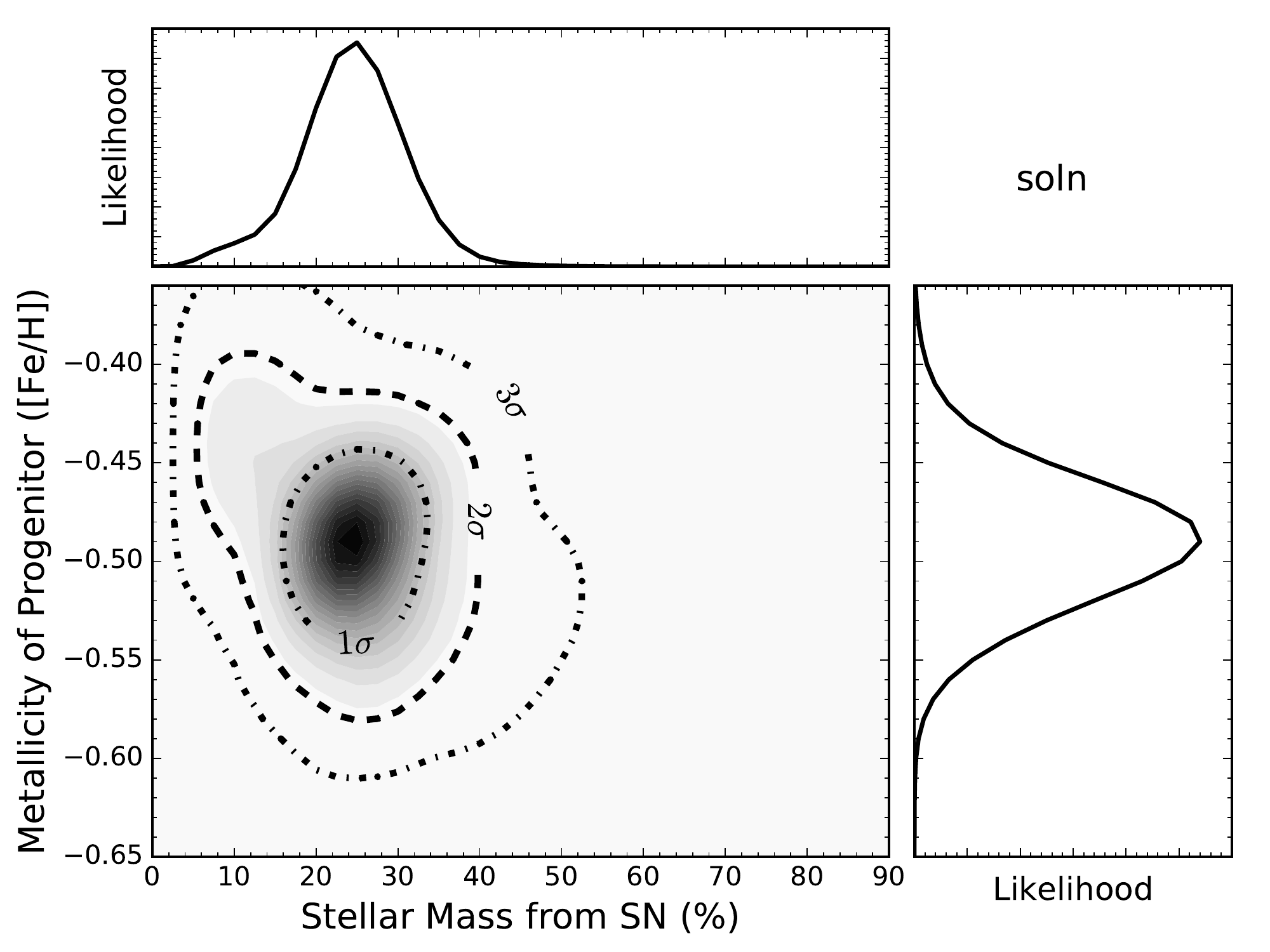}
\caption{Probability distribution of the initial supernova
  progenitor's metallicity and the fractional mass contribution of its
  ejecta to GJ~745 AB.  Confidence intervals indicated by
  1$\sigma$, 2$\sigma$, and 3$\sigma$  enclose 68.3\%, 95.4\%, and
  99.73\% of the total probability, respectively.  The marginalized,
  1D distributions of each parameter are also shown. The best-fit
  values are a $22^{+7}_{-5}$ percent mass contribution and
  metallicity of $-0.48^{+0.03}_{-0.04}$ dex relative to
  Solar. \label{fig:snposterior}}
\end{figure}
\begin{figure}
\plotone{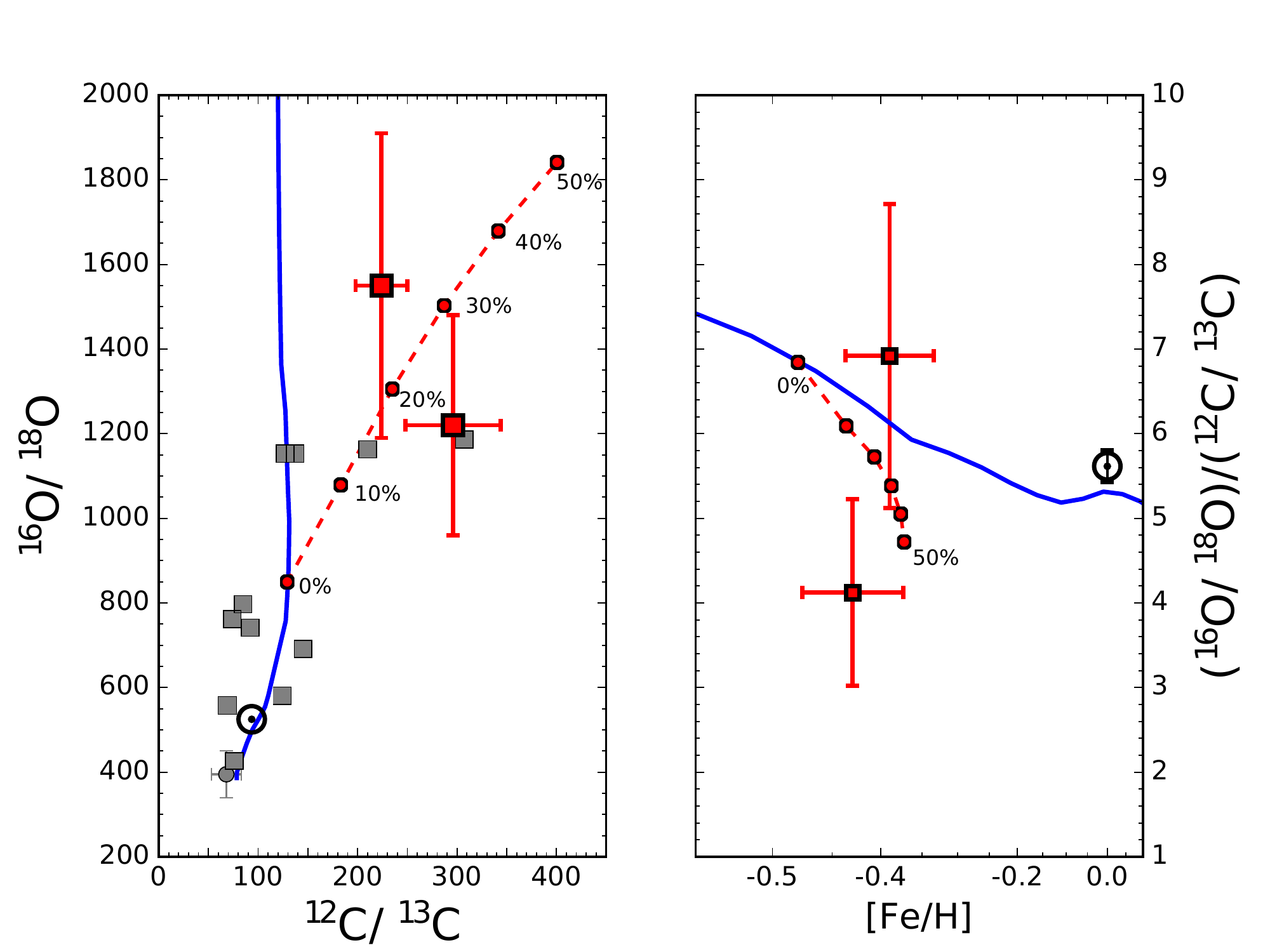}
\caption{Abundance ratios of M dwarfs GJ~745 A and B (red squares) in
  the context of the Sun \citep[dotted circle;][]{lyons:2018}, local
  interstellar medium \citep[gray
    circle;][]{milam:2005,polehampton:2005}, and young stellar objects
  \citep[gray squares;][]{smith:2015}. The M dwarf abundances are
  inconsistent with models of Galactic chemical evolution \citep[blue
    line;][]{kobayashi:2011}, but can be explained by substantial mass
  enrichment from a core-collapse supernova \citep{woosley:1995}.  The
  red dashed line shows an enrichment track for this progenitor star
  at intervals of 10\% stellar mass contribution to the M dwarf
  binary. The best fit is 22\% enrichment by mass from the ejecta of a
  21 $M_\odot$ progenitor with initial metal enhancement of
  $-0.48$~dex. \label{fig:tracks}}
\end{figure}

The remaining, Solar-neightborhood, chemical evolution model can
explain GJ~745AB's \car{12}/\car{13}, \ox{16}/\ox{18}, and [Fe/H] only
through enrichment from CCSN ejecta.  Our best-fit model requires the
injection into a slightly more metal-poor ISM ([Fe/H] =
$-0.48^{+0.03}_{-0.04}$) of ejecta from a CCSN progenitor with mass of
$21 \pm 1 M_\odot$ and an initial metallicity matching that of the
natal ISM, and with $22^{+7}_{-5}\%$ of the M dwarfs' mass consisting
of supernova ejecta (see Figures~\ref{fig:snposterior}
and~\ref{fig:tracks}).  This mass ratio is lower than predictions that
as much as half of the Sun's carbon could have come from supernova
ejecta \citep[][]{clayton:2003}, albeit with a different progenitor
mass and composition. Though such mass fractions may seem large, half
the mass of GJ~745AB ($0.3 M_\odot$) would represent just 0.4\% of the
total ejected mass from such a supernova. If the enrichment came from
multiple supernovae over a short period of time, their individual
contributions would be even less.

This example of strongly-enriched star formation is not unprecedented,
since the isotopic ratios of GJ~745A and B (though not their [Fe/H])
are also consistent with those measured for a handful of young stellar
objects such as IRS~43 \citep{smith:2015}.  If these objects and the
GJ~745AB binary formed in large part from SN ejecta, high-resolution
abundance analyses \citep{souto:2018} should clearly reveal that
formation path, e.g. via enhanced abundances of elements produced by
rapid nucleosynthesis \citep[the ``r-process;''][]{cowan:1991}.
Deeper observations should also enable detection of \co{12}{17}, which
our observations did not have the sensitivity to detect.  The direct
comparison of three O isotopic abundances in a single star would
enable a determination of the level of mass-dependent isotopic
fractionation as has been done for many objects in the Solar System
\citep{clayton:2004}.

Observing the CO fundamental rovibrational band at high resolution has
several clear advantages over similar, past observations of dwarf
stars \citep{pavlenko:2002,tsuji:2016}: (1) we observe the rarer
species in the CO fundamental band, where the cross-sections of these
rare isotopologues are greatest; (2) we resolve individual lines so
blending is not a limitation; and as a result (3) we are sensitive to
much lower abundances of \co{13}{} and \co{}{18}.  The
fundamental-band lines of the CO isotopologues are visible from
spectral types G to L \citep{allard:2014}; modern facilities could
easily measure isotopic abundances in substellar brown dwarfs, and the
next generation of giant ground-based telescopes could extend this
work to the realm of extrasolar planets. The obvious downsides to our
approach are the amount of observing time required per star and the
small number of operational high-resolution, M-band-capable
spectrographs. Nonetheless, such instruments may be poised to open a
new window on stellar isotopic patterns and on our Galaxy's chemical
enrichment history.

\acknowledgments The authors wish to thank many colleagues, as well as
the anonymous referee, for stimulating and thought-provoking
discussions that improved the quality of this work: T. Barman,
Z. Maas, J. Birkby, G. Smith, H. Knutson, C. Pilachowski, I. Roederer,
B. Svoboda, J. Teske, M. Veyette, and E. Newton.  Author
contributions: Conceptualization: IC; Data curation: IC ; Formal
analysis: IC, BF, JV, RF; Funding acquisition: IC; Investigation: IC,
BF, JL, JV, XG, BM; Methodology: IC, JL; Project administration: IC;
Resources: IC, JL; Software: IC, BF, JL, JV; Supervision: IC;
Validation: IC, JL; Visualization: IC; Writing - original draft: IC,
BF, JL, EM; Writing - review and editing: all authors.  Competing
interests: Authors declare no competing interests.  Data and materials
availability: Raw IRTF data will be available for download from the
observatory data archive,
https://irsa.ipac.caltech.edu/applications/irtf/.  The final,
calibrated spectra of GJ745 A and B and the PHOENIX models used in
this work are available in machine-readable format.

Facilities: \facility{IRTF (iSHELL)}

\bibliographystyle{aasjournal}


\begin{longrotatetable}
  \begin{deluxetable}{l c c l}
  \tablecolumns{4} \tablecaption{\label{tab:params} System \& Observational Parameters
  }
  \tablehead{\colhead{Parameter} & \colhead{GJ~745A} & \colhead{GJ~745B} & \colhead{Reference}}
  \startdata
  SpT  & M2V & M2V   &    \\
  $M_{K_s}$ [mag]  & $6.657 \pm 0.021$ &   $6.652 \pm 0.023$   &  \cite{skrutskie:2006,gaia:2018}  \\
  $M_{W2}$ [mag]  & $6.399 \pm 0.022$  & $6.417 \pm 0.031$    &  \cite{cutri:2012,gaia:2018}  \\
  $\varpi$ [mas]  & $113.34 \pm 0.10$ & $113.21 \pm 0.05$   &   \cite{gaia:2018} \\
  $d$ [pc]  & $8.821 \pm 0.008$ & $8.831 \pm 0.004$ & \cite{bailerjones:2018} \\
  $T_\textrm{eff}$ [K] & $3454 \pm 31$ & $3440\pm 31$   & \cite{houdebine:2010,gaidos:2014,mann:2015,newton:2015}   \\
  $R_*$ [$R_\odot$] & $0.32 \pm 0.01$ & $0.33 \pm 0.01$   &  \cite{houdebine:2010,gaidos:2014,mann:2015,newton:2015}  \\
  $M_*$ [$M_\odot$] &$0.31 \pm 0.03$ & $0.31 \pm 0.03$   &  \cite{gaidos:2014,mann:2015,benedict:2016}  \\
  $\log_{10} (L_* / L_\odot)$  & $-1.88 \pm 0.03$ & $-1.91 \pm 0.03$   & This work.   \\
  $v \sin i$ [km~s$^{-1}$] & $<3$ & $<3$   &  \cite{jenkins:2009,houdebine:2010}  \\
  $[Fe/H]$ [dex] & $-0.43 \pm 0.05$ & $-0.39 \pm 0.05$   &  \cite{mann:2015,newton:2015}  \\
  $^{12}$C/H [$10^{-4}$] & $1.04 \pm 0.03$ & $0.995 \pm 0.05$ & This work. \\
  $^{13}$C/H [$10^{-7}$] & $3.5 \pm 0.5$ & $4.5 \pm 0.4$ & This work. \\
  $^{18}$O/H [$10^{-7}$] & $1.7 \pm 0.4$ & $1.3 \pm 0.3$ & This work. \\
  \car{12}/\car{13} &  $296 \pm 45$  &  $224 \pm 26$ & This work \\
  \ox{16}/\ox{18}   & $1220 \pm 260$ & $1550 \pm 360$& This work \\
  \hline
  iSHELL Mode  & M1 & M1   &    \\
  Slit                     & 0.375'' $\times$ 15'' & 0.375'' $\times$ 15''   &    \\
  Guiding filter            & K   & K    &    \\
  Integration Time [sec]   & 14.83 & 14.83   &    \\
  Non-destructive reads & 1 & 1   &    \\
  Co-adds        & 3 & 3   &    \\
  Exposures      & 100 & 102    &    \\
  UT Times &  08:19 -- 09:47  & 10:35 -- 12:06   &    \\
  Slit Position Angle & -84.9 -- -88.6$^o$ & -108.1 -- $+99.5^o$   &    \\
  Airmass range            & 1.22 - 1.03 & 1.00 - 1.05   &    \\
  A0V airmass range        & \multicolumn{2}{c}{1.05--1.01} 
\enddata
\end{deluxetable}    
\end{longrotatetable}

\begin{deluxetable}{rlllllllll}
\tabletypesize{\small}\tablecaption{CO Lines Used In This Work}\tablehead{
  \colhead{$\lambda_0$} & & \multicolumn{4}{c}{Transition Details} & \multicolumn{4}{c}{Continuum Regions\tablenotemark{a}}   \\
\colhead{[\AA]} &  \colhead{species} & \colhead{$\nu'$} & \colhead{$\nu''$} & \colhead{branch} & \colhead{$J$} & \colhead{$L_L$} & \colhead{$L_R$} & \colhead{$R_L$} & \colhead{$R_R$}  
}\startdata
23374.128 & \co{12}{16} & 2 & 0 & R & 4  & -3.1 & -2.1 & 1.6 & 2.5 \\
23393.233 & \co{12}{16} & 2 & 0 & R & 3  & -1.9 & -1.3 & 1.7 & 2.2 \\
23412.752 & \co{12}{16} & 2 & 0 & R & 2  & -4.5 & -2.5 & 1.3 & 1.7 \\
23432.695 & \co{12}{16} & 2 & 0 & R & 1  & -1.5 & -0.6 & 0.6 & 1.9 \\
23727.167 & \co{12}{16} & 3 & 1 & R & 1  & -5.0 & -4.0 & 1.8 & 2.5 \\ \hline
46116.476 & \co{13}{16} & 1 & 0 & R & 21 & -9.0 & -2.0 & 5.0 & 7.0 \\
46178.775 & \co{13}{16} & 1 & 0 & R & 20 & -8.0 & -5.5 & 2.0 & 5.5 \\
46205.730 & \co{13}{16} & 2 & 1 & R & 29 & -0.9 & -0.65& 1.0 & 2.5 \\
46241.999 & \co{13}{16} & 1 & 0 & R & 19 & -5.0 & -2.5 & 4.0 & 6.0 \\
46317.954 & \co{13}{16} & 2 & 1 & R & 27 & -4.5 & -1.5 & 4.2 & 6.5 \\
46404.576 & \co{13}{16} & 3 & 2 & R & 36 & -4.0 & -1.5 & 1.0 & 2.0 \\
46433.864 & \co{13}{16} & 2 & 1 & R & 25 & -5.0 & -1.5 & 5.0 & 9.0 \\
46556.783 & \co{13}{16} & 3 & 2 & R & 33 & -8.0 & -4.5 & 1.5 & 3.8 \\
46641.064 & \co{13}{16} & 1 & 0 & R & 13 & -3.0 & -1.3 & 1.4 & 2.5 \\
46710.901 & \co{13}{16} & 1 & 0 & R & 21 & -4.0 & -3.0 & 2.0 & 4.0 \\
46717.315 & \co{13}{16} & 3 & 2 & R & 30 & -3.5 & -1.2 & 2.0 & 5.0 \\
46926.227 & \co{13}{16} & 1 & 0 & R & 9  & -4.0 & -1.5 & 1.5 & 3.0 \\
46935.010 & \co{13}{16} & 2 & 1 & R & 17 & -3.5 & -1.8 & 2.0 & 5.5 \\ \hline
46030.116 & \co{12}{18} & 2 & 1 & R & 34 & -3.5 & -1.0 & 1.0 & 2.5 \\
46133.337 & \co{12}{18} & 2 & 1 & R & 32 & -4.5 & -1.3 & 3.6 & 4.6 \\
46144.648 & \co{12}{18} & 1 & 0 & R & 22 & -1.8 & -0.8 & 0.8 & 1.4 \\ 
46186.304 & \co{12}{18} & 2 & 1 & R & 31 & -4.5 & -1.5 & 1.0 & 2.5 \\
46240.182 & \co{12}{18} & 2 & 1 & R & 30 & -2.1 & -1.1 & 0.5 & 4.0 \\
46594.024 & \co{12}{18} & 1 & 0 & R & 15 & -2.6 & -1.0 & 0.8 & 1.7 \\
46704.254 & \co{12}{18} & 2 & 1 & R & 22 & -4.0 & -1.0 & 1.0 & 2.5 \\
46871.567 & \co{12}{18} & 1 & 0 & R & 11 & -1.9 & -0.9 & 1.0 & 4.0 \\
46893.682 & \co{12}{18} & 2 & 1 & R & 19 & -4.0 & -1.0 & 1.0 & 2.5 \\
47016.125 & \co{12}{18} & 1 & 0 & R & 9  & -2.1 & -1.0 & 1.0 & 2.1 \\
47024.717 & \co{12}{18} & 2 & 1 & R & 17 & -4.0 & -0.5 & 0.5 & 4.0 
\enddata\label{tab:lines}
\tablenotetext{a}{Left ($L$) and right ($R$) edges of the left- and right-hand regions used for linear continuum normalization; all values are relative to $\lambda_0$, in \AA.}
\end{deluxetable}

\end{document}